\documentstyle [psfig]{laa}
\newcommand {\etl} {\mbox{et al.}}

\newcommand {\lii}  {Li\,{\sc i}}

\begin{document}

\newcommand {\Tf} {T_{\rm eff}}
\newcommand {\lgg}{\log g}
\newcommand {\SB} {$S_{l}/B_{\nu}$}
\def\lN{\log N({\rm Li})}

\thesaurus{07 (02.12.1; 02.12.3; 08.01.1; 08.12.1)}
\title {Theoretical LTE \& non-LTE curves of growth for Li\,{\sc I}  lines in G-M dwarfs and subgiants}
\author{Ya. V. Pavlenko\inst{1} \and A.~Magazz\`u\inst{2}}
\offprints {Ya.V. Pavlenko}
\institute{The Main Astronomical Observatory of Academy of Sciences of Ukraine,
Golosiiv, 252650 Kyiv-22, Ukraine
\and Osservatorio Astrofisico di Catania, Citt\`a Universitaria, I-95125 Catania, Italy}
\date {Received date; accepted date}
\maketitle
\begin{abstract} 
Detailed simultaneous radiative transfer-statistical equilibrium
calculations have been performed for lithium in atmospheres of solar
metallicity G-M dwarfs and subgiants. Model atmospheres of stars with
$3500 \le \Tf \le 6000$~K, $3.0 \le \lgg \le 4.5$ have been
considered.  The behaviour of  level populations, source functions and
curves of growth for the lithium doublet at $\lambda 670.8$~nm and two
subordinate lines of astrophysical interest ($\lambda 610.3$~nm and
$\lambda 812.6$~nm) are discussed. For the coolest atmospheres of our
grid we find that the  dependence of non-LTE equivalent widths and
profiles on gravity is  very weak. The computed LTE and non-LTE curves
of growth are given in tabular form.
\keywords{Line: formation -- Line: profiles -- Stars:  abundances -- Stars: late type}
\end{abstract}

\section {Introduction}

The importance of lithium in many fields of astrophysics has been
widely recognized during the last decades. The proceedings of a recent
ESO/EIPC workshop (Crane 1995) describe the status of the present
knowledge on this subject. The abundance of lithium is usually
determined from the resonance doublet at $\lambda 670.8$~nm. The
equivalent width of this feature, quite small in general,  can reach
values up to  hundreds of mA in  spectra of young pre-main sequence
objects, where the doublet can get easily saturated. We restricted our
study to stars with  $\Tf \ge 3500$~K\@. Most lithium in the atmosphere of
these stars is in the form of Li\,{\sc ii} ions. Therefore, the amount
of neutral lithium can be sensitive to non-LTE (non local thermodynamic
equilibrium) effects, notably overionization, leading to non-LTE
effects on abundance determinations, even for weak lines. Strong lines
in addition will have non-LTE affecting the source function. Studies
of the non-LTE  formation of \lii\ lines are required in order to get
accurate abundances from the observed equivalent widths.

We have presented our results on the non-LTE formation of lithium lines
in the atmosphere of cool dwarfs and subgiants in several  papers
(Magazz\`u \etl\ 1992, Mart\'\i n \etl\ 1994, Pavlenko 1995, Pavlenko
\etl\ 1995), aimed mainly to determine the lithium abundance in the
atmosphere of pre-main sequence objects. Recently, Carlsson
\etl\ (1994) published results of extensive non-LTE computations for
lithium for a grid of model atmospheres with $\Tf$ between 4500 and
7500~K.  They provide numerical corrections to get  non-LTE Li
abundances from LTE (local thermodynamic equilibrium) curves of
growth.

In this work we present new results of  non-LTE studies concerning the
\lii\ resonance  line at $\lambda 670.8$~nm and also the subordinate
lines at $\lambda 610.3$~nm and $\lambda 812.6$~nm, whose astrophysical
interest has been  recently recognized (Duncan 1991, Magazz\`u
\etl\ 1995). The curves of growth presented here for the $\lambda
670.8$~nm line  are a refinement of those used by  Mart\'\i n
\etl\ (1994).

\section {The Procedure}

To solve the system of statistical balance equations and radiative
transfer equations  (the non-LTE problem) we followed the modification
of the linearization method proposed by Auer and Heasley (1976).  Our
computation procedure of the non-LTE problem solution was described
extensively in  Pavlenko \etl\ (1995). Some details  are given below:

\begin {itemize}

\item we used a 20-level model of the \lii\ atom;

\item the 70 radiative and all possible collisional transitions were
included into the rate matrix;

\item the ionization-dissociation equilibrium was computed for 98 atoms
and molecules including 6 lithium-containing molecules;

\item  the multiplet radiative transitions were replaced by single
transitions, even though absorption coefficient profiles were computed
taking into account the multiplet structure;

\item the opacity due to atomic lines from Kurucz (1979) was taken into
account.  Additionally, we included the  absorption due to   molecular
bands in JOLA approximation, which gives good results for saturated
bands (Tsuji 1994).  Twenty-one molecular bands have been taken into
account. For unsaturated molecular bands (in the case of stars with
$\Tf > 4000$~K) results are not so good, but in this case the
visible and violet parts of the spectrum are dominated by absorption
due to atomic lines. All these opacities were treated in  LTE;

\item we used the correction factor $E = 2$ for the Uns\"old
approximation of van der Waals damping for lithium lines (see
Section~2.5  in Pavlenko \etl\ 1995);
 
\item we excluded from our computations the inelastic collisions of
\lii\ atoms with hydrogen neutral atoms, due to the strong criticism
raised (see Lambert 1993, Carlsson \etl\ 1994) on Drawin's (1968, 1969)
formulae and their modifications by Steenbock and Holveger (1984).

\end {itemize}

The main difference between this work and  computations by  Pavlenko
\etl\ (1995) is the model atmospheres. Here we use models calculated
with ATLAS9 and taken from Kurucz (1993) CD-ROM~13.  We consider only
solar metallicity models, with microturbulence of $2~{\rm km~s}^{-1}$.
Any  contribution from $^6$Li to the equivalent widths of the lines in
study has been neglected.

\section{Results}
 
We treat here the range $3500 \le \Tf \le 6000$~K, which includes
atmospheres where the  molecular absorption is relevant.  In such
atmospheres the thermalization processes  increase due to the strong
absorption by atomic and molecular lines in the frequencies of
\lii\ resonance and subordinate lines.  These processes are currently
studied and  will be discussed in forthcoming papers.  Here we give a
few results, relevant for understanding the nature of non-LTE effects
in \lii\ lines in cool  atmospheres.  For this purpose it is useful
to  consider the dependence  of departure coefficients and source
functions of the absorption lines on the optical depth.

\subsection {The departure coefficients}

\begin {figure}
\psfig {file=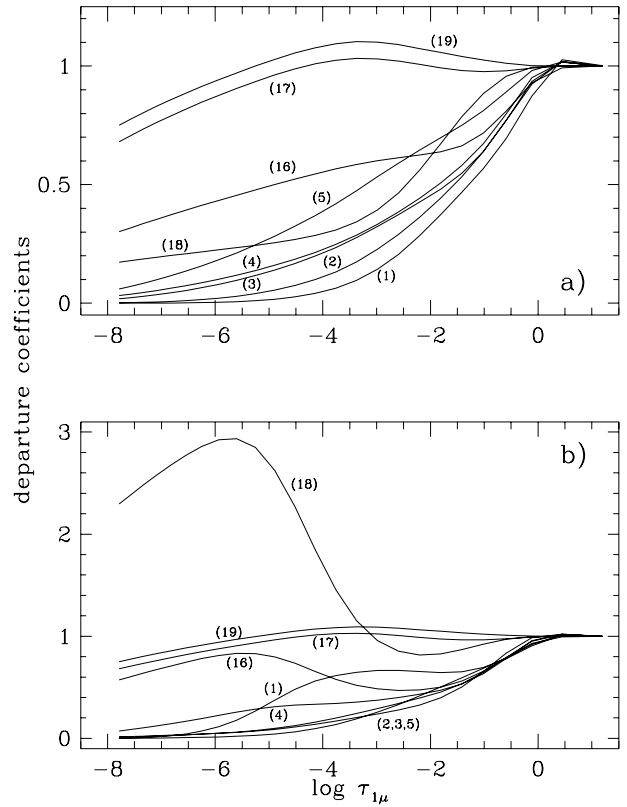,height=11.3cm}
\caption [] {Departure coefficients of \lii\ levels in atmosphere of a 4000/3.0
        subgiant with  a lithium abundance $\lN = 0.5$ (a) and $\lN = 3.5$ (b)}
\label {dep}
\end {figure}

In  Fig.~\ref{dep} we present the departure coefficients of lithium
levels $b_i = n_i / n^{*}_i$ versus $\tau_{1\mu}$, the continuum
optical depth at $1~\mu$m, in the atmosphere of a dwarf 4000/3.0,
i.e.\ with $\Tf = 4000$\,K, $\lgg = 3.0$. Here $n^{*}_i$ and $n_i$ are
the LTE and non-LTE populations of the lithium levels, respectively.
The  model atom used was described in detail by Pavlenko (1994). We
present results of computations for abundances\footnote{Throughout this
paper $\lN$ is given in the scale $\log N({\rm H}) =12$} $\lN = 0.5$
and $\lN = 3.5$, i.e. in the case of weak and strong lithium resonance
doublet, respectively.

We find that:

\begin {itemize}

\item the behaviour of the departure coefficients depends on the
strength of \lii\ resonance doublet, i.e.\ on the  lithium abundance
(see also Magazz\`u \etl\ 1992);

\item properly speaking, we should use the  definition ``overionization
of lithium" to describe only the relevant process of formation of the
statistical equilibrium of lithium. In fact, the statistical
equilibrium of lithium depends on a whole set of  radiative and
collisional transitions (see Carlsson \etl\ 1994).  Still, in the
formation region  of  weak lithium $\lambda 670.8$\,nm lines ($\tau
\equiv \tau_{1\mu} \sim 10^{-2}$) overionization dominates. As a result
we have here $b_{1} < b_{2} < 1.0$ (see also Steenbock \& Holweger
1984);

\item  in the  case of saturated lithium resonance lines we have balance of the
radiative transitions (2s-2p) in a large part of the stellar
atmosphere. The population of the first lithium level is formed by
chains of transitions from upper levels and continuum. In the formation
region of strong resonance lithium lines  ($\tau \sim 10^{-5}$) we
have  $b_{1} > b_{2}$.  In the outer part of stellar atmospheres the
first level  may be even overpopulated with respect to LTE (see also
Pavlenko 1992, Carlsson et al. 1994);

\item in addition, the departure coefficients of upper levels reach
values greater than 1 in some regions of the atmosphere. These levels
are  more tied with continuum than with lower levels.

\end {itemize}

\subsection  {The source function}

\begin {figure}[t]
\psfig {file=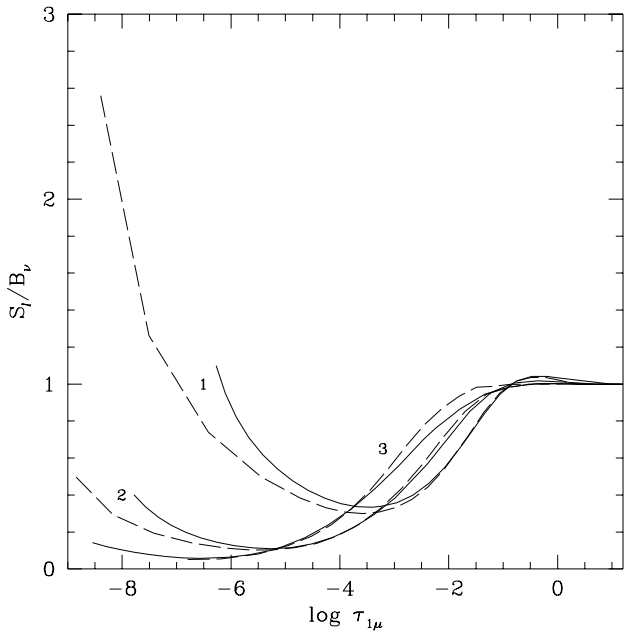,height=8.8cm}
\caption []{Ratio $S_{l}/B_{\nu}$ for the \lii\ resonance doublet in
atmospheres  5000/4.5 and 5000/3.0 (1), 4000/4.5 and 4000/3.0 (2),
3500/4.5 and 3500/3.0 (3).  Solid lines correspond to $\lgg = 3.0$,
dashed ones to $\lgg = 4.5$, with a lithium abundance $\lN = 3.5$}
\label{rat}
\end {figure}

The nature and intensity of the non-LTE effects depend on \SB, i.e.\ the
ratio of the source function to the Planck function at the formation
depth of a line. Here we study such a dependence for the lines
considered in this work. In  Fig.~\ref{rat} we present the ratio
\SB\ computed for the \lii\ $\lambda 670.8$~nm resonance doublet 
plotted against $\tau_{1\mu}$ for several model atmospheres with
different effective temperatures and gravities.

We find that in the $\lambda 670.8$~nm line formation region:

\begin {itemize}

\item the ratio  \SB\ is lower for stars with lower $\Tf$;

\item the ratio is lower in atmospheres of lower luminosity stars;

\item the differences in \SB\ computed for $\lgg = 3.0$ and 4.5
 are  very small for $\Tf = 3500$~K.

\end {itemize}

When the effective temperature  decreases the lithium resonance lines
(for a given lithium abundance) become stronger, so their thermalization
depths shift towards the outer boundary of the atmosphere.
Note that the dependence of thermalization processes on the gravity is
less pronounced.

\begin {figure}
\psfig {file=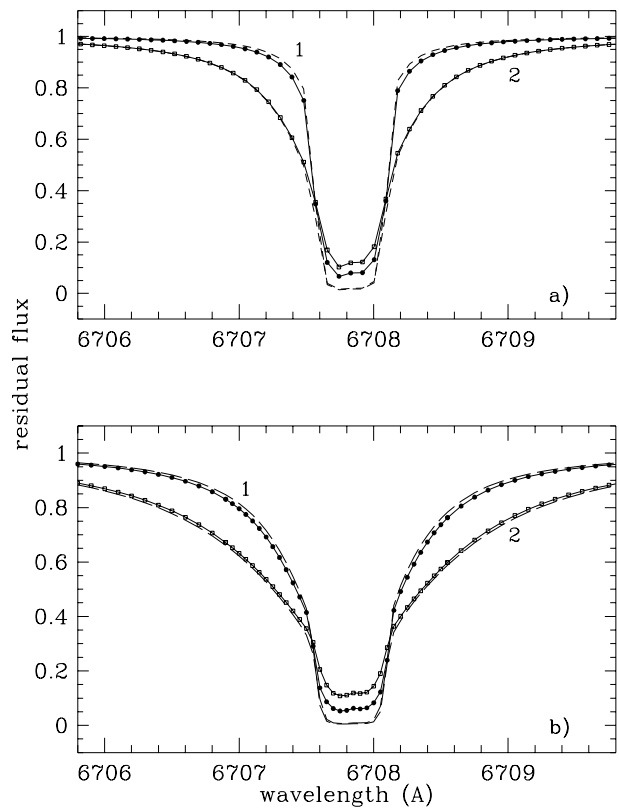,height=11.3cm}
\caption [] {The LTE and non-LTE profiles of the \lii\ resonance
doublet computed for a) 4000/3.0 (1) and 4000/4.5 (2) model
atmospheres; b) 3500/3.0 (1) and 3500/4.5 (2) model atmospheres.
Solid   and dashed lines show LTE and non-LTE results, respectively.
Dots indicate $\lgg = 3.0$ and open squares  $\lgg = 4.5$. Note how
the cores of non-LTE profiles do not differ for different $\lgg$'s}
\label {pro}
\end {figure}

\begin {figure}
\psfig {file=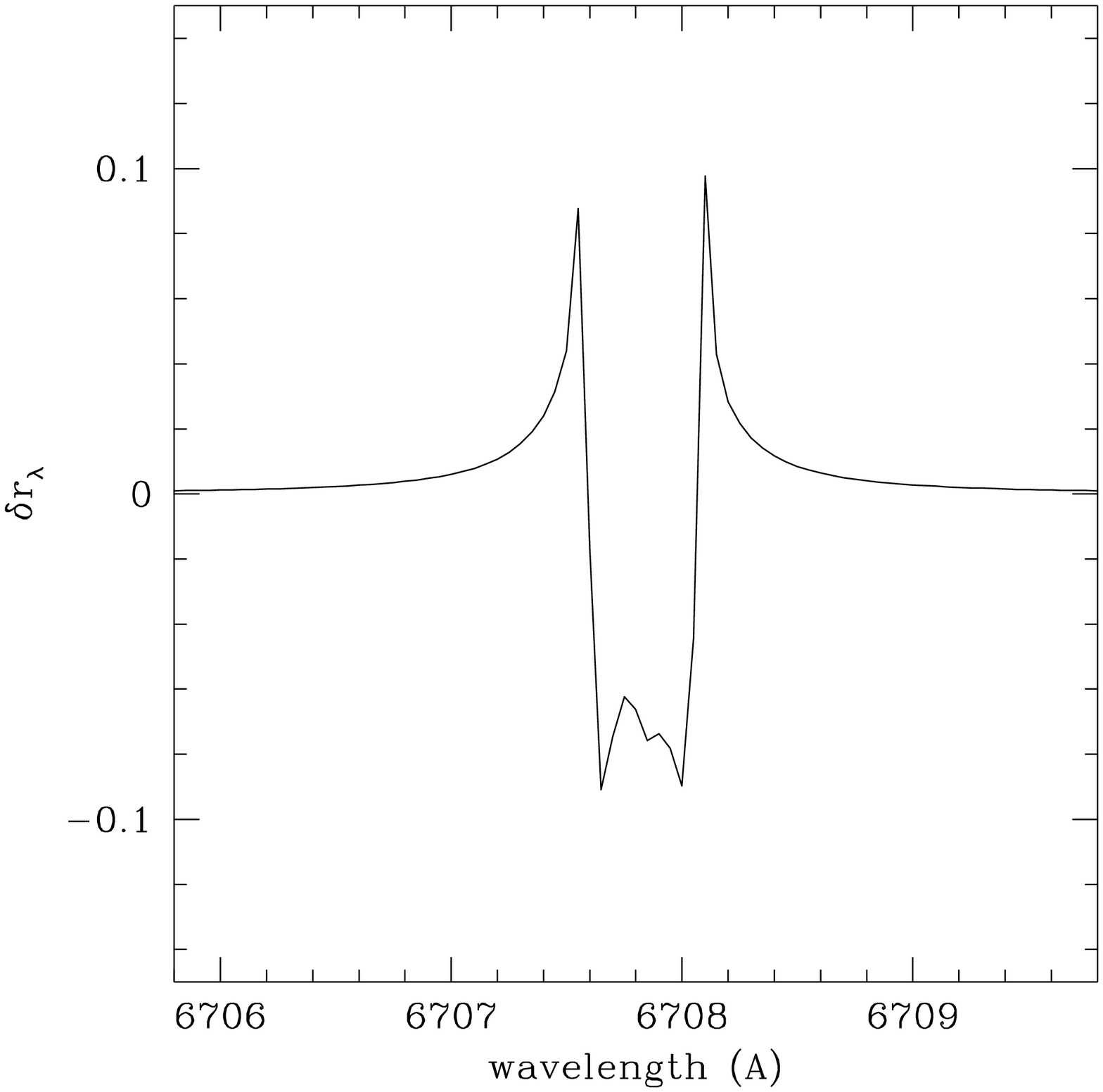,height=8.8cm}
\caption [] {The difference between a LTE and a non-LTE \lii\ resonance
doublet of the same equivalent width ($W_{\lambda} = 30.1$ pm). Model
atmosphere is 4000/3.0}
\label{com}
\end {figure}

In the case of the subordinate lines we find that in the outer part of the
atmosphere $S_{l}/B_{\nu} > 1.0$, while in the formation region of these  lines
($\tau \sim 0.01$) is $S_{l}/B_{\nu} \sim  1.0$.  The ratio \SB\ for the
$\lambda 610.3$~nm line is greater than for the $\lambda 812.6$ nm
line, because in  the frequencies of the former line $\partial B/\partial
\tau$ is greater.

\subsection{The profile of the resonance doublet}

Here we compare the LTE and non-LTE profiles of the resonance doublet.
The absorption  doublet is saturated in the case of stellar atmospheres
with $\Tf \le 4000 $ and  $\lN > 2.0$. Thus,  to describe the radiation
transfer in the line core frequencies we extrapolated the Kurucz model
atmospheres  up to levels where line cores are formed, as described in
Pavlenko (1984) and Magazz\`u \etl\ (1992). In  Fig.~\ref{pro} we show
a few profiles computed for lithium abundance $\log N({\rm Li}) =
3.5$.  We see that non-LTE cores are deeper in comparison with LTE, as
found in Magazz\`u \etl\ (1992); see also Pavlenko (1992) and Carlsson
\etl\ (1994).  The \lii\ resonance line computed for model atmospheres
with larger $\lgg$ shows  stronger wings, both in LTE and non-LTE, due
to the pronounced damping. However, we see very small differences in
the non-LTE cores computed for model atmospheres with different
$\lgg$'s. In the case of unsaturated \lii\ resonance line the non-LTE
profiles are shallower than the LTE ones (see Fig.~7 in Pavlenko
\etl\ 1995).

The shape of the Li doublet becomes extremely important in some
astrophysical contexts (e.g.\ Li isotopic ratio determinations). The
difference between a LTE and a non-LTE  line profile, calculated for
the same equivalent width and for a model atmosphere 4000/3.0, is
shown  in Fig.~\ref{com}.\@  The largest differences are found in the
core and core-wing transition part of the line.  Note that these
differences are negative in the core and positive in the wings.

\begin{table*}
\caption[]{Curves of growth of the $\lambda 670.8$ nm line}
\label{tab1}
\begin{tabular} {lrrrrrrr}
\hline\hline
\noalign{\smallskip}
           &$\log N({\rm Li})$ & \multicolumn {6}{c}{$W_\lambda~({\rm mA})$}      \\ \cline {3-8}
\multicolumn{2}{c} { } & \multicolumn{2}{c} {$\log g = 3.0$} & \multicolumn{2}{c}
{$\log g = 3.5$} &\multicolumn{2}{c} {$\log g = 4.5$} \\ 
     & & LTE   &  NLTE   &  LTE  &   NLTE   &    LTE    &  NLTE  \\
\noalign{\smallskip}
 \hline
\noalign{\smallskip}
& $-1.5$ &     9.3 &     4.8&     7.6 &     4.7     &      5.6 &     3.7   \\
& $-1.0$ &    27.9 &    14.8&    23.1 &    14.4     &     16.9 &    14.7 \\
& $-0.5$ &    75.5 &    42.6&    63.9 &    42.3     &     48.1 &    42.4  \\
&  0.0 &   163.3 &   106.6&   144.3 &   105.2       &    116.0 &   106.4 \\
&  0.5 &   258.1 &   208.9&   240.8 &   210.2       &    211.6 &   210.8  \\
$T_{\rm eff} = 3500K$&  1.0 &   330.6 &   309.2& 317.3 &   310.3 &    298.6 &   326.3 \\
&  1.5 &   394.0 &   390.1&   385.3 &   408.9       &    384.6 &   440.8 \\
&  2.0 &   467.8 &   471.2&   469.9 &   492.5       &    509.6 &   589.8  \\
&  2.5 &   584.8 &   584.5&   889.9 &   929.7       &    735.9 &   839.7  \\
&  3.0 &   815.5 &   798.4&   889.3 &   914.0       &   1166.1 &  1287.7  \\
&  3.5 &  1271.5 &  1219.1&  1427.4 &  1454.1       &   1957.1 &  2087.9  \\
&  4.0 &  2121.7 &  1991.9&  2405.6 &  2371.1       &   3335.8 &  3466.8    \\
\noalign{\smallskip}
\hline
\noalign{\smallskip}
&  $-0.5$ &    16.1 &     7.4 &      15.7&      7.7&    11.5 &     7.9 \\
&   0.0 &    46.5 &    22.6 &      45.2&     23.4&    34.0 &    23.8 \\
&   0.5 &   114.4 &    62.9 &     111.8&     65.0&    88.6 &    66.1 \\
&   1.0 &   209.7 &   147.1 &     206.5&    150.8&   179.6 &   153.5  \\
&   1.5 &   289.1 &   262.3 &     286.8&    267.3&   270.5 &   274.1 \\
$T_{\rm eff} = 4000K$ & 2.0 & 353.9 & 363.1 &     355.3&    370.1&   354.4 &   393.3  \\
&   2.5 &   423.1 &   444.0 &     436.0&    461.8&   469.0 &   525.4 \\
&   3.0 &   513.8 &   530.0 &     559.3&    575.2&   669.5 &   733.3 \\
&   3.5 &   676.0 &   669.6 &     793.1&    784.4&  1049.8 &  1115.9  \\
&   4.0 &   995.8 &   944.9 &    1246.9&   1188.5&  1751.1 &  1800.3  \\ 
\noalign{\smallskip}
\hline
\noalign{\smallskip}
&   0.0 &     8.9 &     4.2 &      8.9 &     4.4  &     8.8 &     4.9 \\
&   0.5 &    26.8 &    13.1 &     26.8 &    13.7  &    26.4 &    15.3 \\
&   1.0 &    73.1 &    38.3 &     73.0 &    40.0  &    72.3 &    44.5 \\
&   1.5 &   159.9 &    99.5 &    159.2 &   103.3  &   158.5 &   113.4  \\
&   2.0 &   255.2 &   205.0 &    253.5 &   210.6  &    254.7  &   227.7  \\
$T_{\rm eff} = 4500K$ & 2.5 &  331.3 & 318.6 &    329.6 &   325.0  & 340.3 &   350.8  \\
&   3.0 &   401.3 &   407.4 &    403.6 &   417.0  &   443.0 &   468.9 \\
&   3.5 &   474.0 &   482.3 &    490.6 &   503.6  &   604.3 &   625.5 \\
&   4.0 &   572.3 &   569.2 &    625.7 &   623.6  &   889.4 &   893.4  \\ 
\noalign{\smallskip}
\hline
\noalign{\smallskip}
& 1.0& 17.2&  10.6&  17.8&  11.1 &  18.7&  12.4  \\ 
& 1.5& 48.9&  31.6&  50.5&  32.8 &  52.8&  36.4  \\
& 2.0& 117.4&  84.3&  120.3&  87.0 &  124.9&  95.5  \\
& 2.5& 209.7&  182.3&  212.6&  186.7 &  218.5&  201.1   \\
$T_{\rm eff} = 5000K$& 3.0& 288.0&  297.9&  290.5&  302.9 &  299.5&  321.7  \\
& 3.5& 355.0&  392.4&  358.7&  398.2 &  379.1&  426.2   \\
& 4.0& 420.4&  466.1&  429.1&  475.9 &  478.2&  529.1   \\
\noalign{\smallskip}
\hline
\noalign{\smallskip}
 & 1.5& 14.4&  11.8&   15.1&  12.1&   16.0&  12.9  \\
 & 2.0& 41.4&  34.7&   43.3&  35.5&   45.7&  37.9  \\
 & 2.5& 101.5&  91.1&   105.2&  92.9&   110.3&  98.6   \\
$T_{\rm eff} = 5500K$& 3.0& 186.6&  191.8&   191.1&  194.8&   198.8&  204.6 \\
 & 3.5& 261.4&  306.5&   266.2&  309.9&   277.2&  323.1  \\ 
 & 4.0& 324.7&  399.7&   330.8&  403.9&   350.8&  423.3 \\ 
\noalign{\smallskip}
\hline
\noalign{\smallskip}
& 2.0&  15.0&  15.1& 16.3&  15.7& 17.0&  16.4    \\
& 2.5&  42.8&  43.6& 46.0&  45.2& 47.9&  47.2     \\
$T_{\rm eff} = 6000K$& 3.0&  102.5&  109.8& 108.3&  113.1& 112.4&  117.7    \\
& 3.5&  183.2&  217.0& 189.8&  221.7& 195.9&  229.3     \\
& 4.0&  253.2&  328.9& 260.0&  333.5& 269.4&  343.4    \\
\noalign{\smallskip}
\hline \hline
\end{tabular}
\end{table*}

\begin {figure} 
\psfig {file=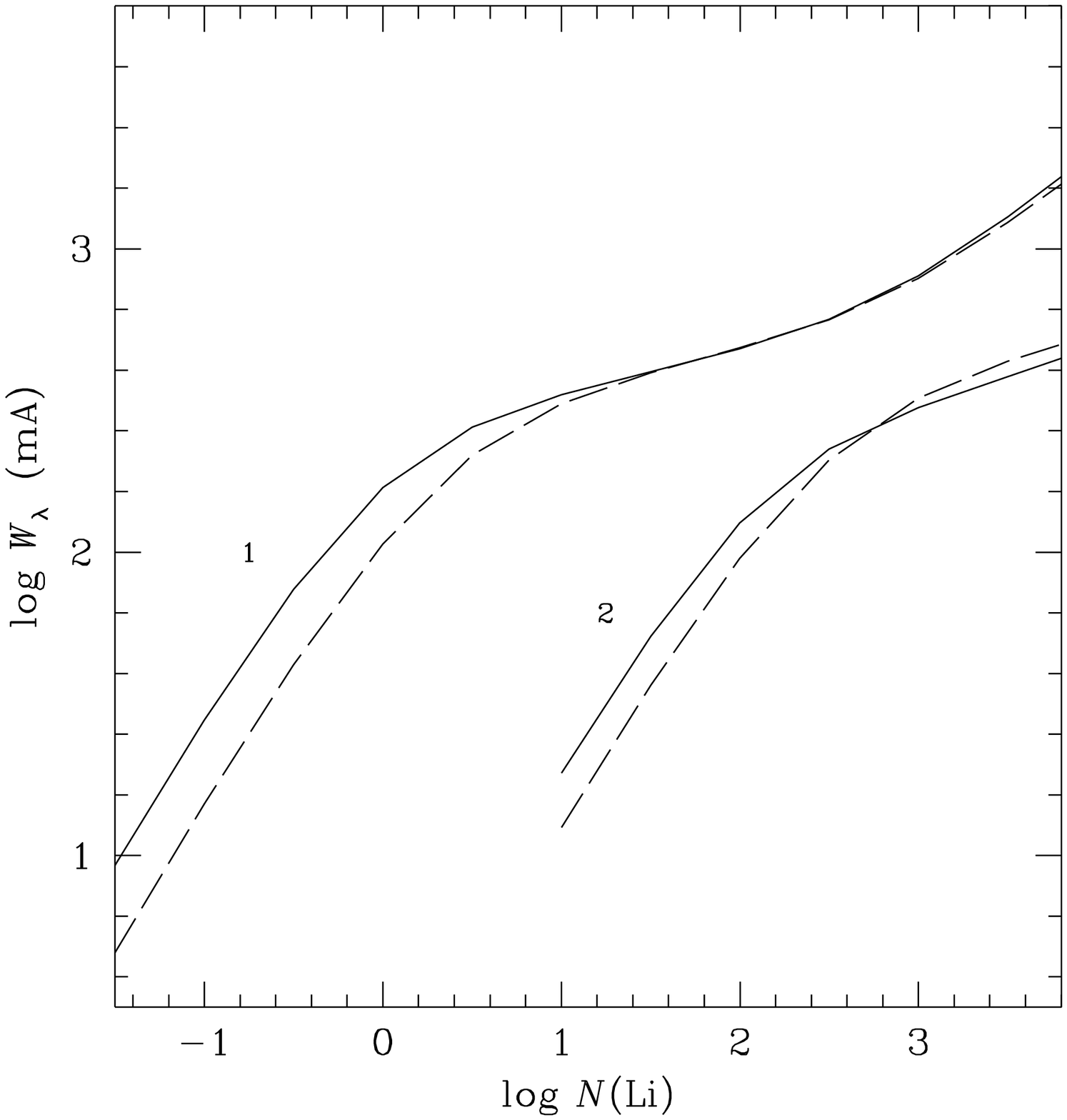,height=8.8cm}
\caption [] {The LTE and non-LTE curves of growth of the
\lii\ resonance doublet computed for  3500/3.0 (1) and 5000/4.5 (2).
Solid and dashed lines show the LTE and non-LTE results, respectively}
\label{co67}
\end {figure}

\subsection {The curves of growth}

\subsubsection {Resonance doublet}

The  LTE and non-LTE curves of growth of the resonance doublet,
computed for model atmospheres of several dwarfs and subgiants are
presented in Table~\ref{tab1}\@. In Fig.~\ref{co67} we plot some of
these curves, corresponding to the models 3500/3.0 and 5000/4.5.\@ We
see that non-LTE abundance corrections   $\Delta \log N({\rm Li}) =
\log N^{\rm NLTE}({\rm Li}) - \log N^{\rm LTE}({\rm Li})$ are positive
for unsaturated lines formed in the atmospheres of dwarfs and
subgiants.  As noted before, in that case the influence of
overionization of lithium  dominates.  The formation region of the
non-LTE doublet shifts (with respect to LTE) towards deeper layers,
where $S_{l}(\tau_{\rm nlte}=1) > B_{\nu}(\tau_{\rm lte}=1)$, being
$\tau$ the optical depth at the frequency of the resonance line. A
stronger doublet is formed where $S_{l}(\tau_{\rm  nlte}=1) <
B_{\nu}(\tau_{\rm lte}=1)$. As a result, the sign of $\Delta \log
N({\rm Li})$ changes (Magazz\`u \etl\ 1992). Again, in the case of the
subgiant 4000/3.0, when the doublet is extremely strong $\Delta \log
N({\rm Li})$ changes its sign once more. The corresponding   line
formation region moves towards the outer boundary of the atmosphere
where the overionization increases with height and wins over the source
function effect. This tendency can  be seen also in the top panels of
Figs.~16 and 17 in Carlsson \etl\ (1994) and in Fig.~4 in Pavlenko
\etl\ (1995).

Another result seems to be more interesting:  our computations show
that non-LTE curves of growth computed for dwarfs and subgiants with
$\Tf \le 4000$~K do not differ significantly in the case of unsaturated
lines. In other words, whereas in  LTE the equivalent width of the
(unsaturated) \lii\ doublet is a function of $T_{\rm eff}$, $\log g$, and
$N({\rm Li})$,  non-LTE computations give  $W_{\lambda}$   as a
function of only $T_{\rm eff}$ and $N({\rm Li})$.  This is a very
useful result, because the determination of the gravity usually is more
difficult than the assignment of an effective temperature.

\begin {figure}
\psfig {file=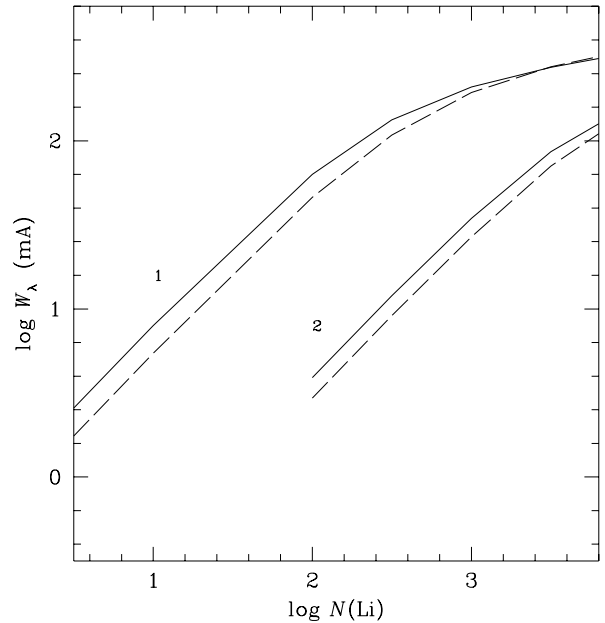,height=8.8cm}
\caption [] {The same as in Fig.~\ref{co67},  but for subordinate line 
at $\lambda 610.3$~nm}
\label {co61}
\end {figure}

\subsubsection  {Subordinate lines}

In Tables~2,\,3 and Figs.~6,\,7  we show   the case of the subordinate
\lii\ lines at $\lambda 610.3$ and $\lambda 812.6$\,nm.  These lines
are less intense than the resonance doublet and form in deeper parts of
the atmosphere (Pavlenko \etl\ 1995). In spite of this, for these
unsaturated lines we have  also determined that the equivalent width in non-LTE
practically does not depend on gravity for the model atmospheres with
$T_{\rm eff} \le 4000$K\@.

\begin{table*}
\caption[]{Curves of growth of the $\lambda 610.3$ nm line}
\label{tab3}
\begin{tabular}{lrrrrrrr}
\hline
\hline
\noalign{\smallskip}
           &$\log N({\rm Li})$ & \multicolumn {6}{c}{$W_\lambda~({\rm mA})$}      \\ \cline {3-8}
\multicolumn{2}{c} { } & \multicolumn{2}{c} {$\log g = 3.0$} & \multicolumn{2}{c}
{$\log g = 3.5$} &\multicolumn{2}{c} {$\log g = 4.5$} \\ 
     & & LTE   &  NLTE   &  LTE  &   NLTE   &    LTE    &  NLTE  \\
\noalign{\smallskip}
\hline
\noalign{\smallskip}
               &   0.5&  2.6 &  1.7&   2.2&  1.7 &   1.7 &  1.7 \\
               &   1.0&  8.0 &  5.4&   6.7&  5.4 &   5.3 &  5.3 \\
               &   1.5& 23.7 & 16.4  &   20.2&  16.2 &   15.9 &  16.1 \\
$T_{\rm eff} = 3500K$&   2.0&  63.2&  46.1&   54.8&  45.2 &   44.2&  44.8 \\
               &   2.5&  133.4&  108.1&   120.1&  105.9 &   101.7&  105.0 \\
               &   3.0&  209.4&  193.7&   197.2&  190.9 &   179.8&  192.7 \\
               &   3.5&  274.1&  275.8&   265.6&  275.8 &   258.4&  288.0 \\
               &   4.0&  334.8&  348.3&   331.6&  354.9 &   342.7&  389.0 \\
\noalign{\smallskip}
\hline
\noalign{\smallskip}
               &   1.0&  3.0&  1.8&  3.0 &  1.9 &  2.3&  1.9 \\
               &   1.5&  9.3&  5.6&  9.2&  5.8 &  7.1&  5.8   \\
               &   2.0&  27.6&  16.9&  27.3&  17.5 &  21.4&  17.6 \\
               &   2.5&  72.2&  47.2&  71.7&  48.7 &  58.1&  49.1 \\
$T_{\rm eff} = 4000K$&   3.0&  146.9&  111.1&  146.2&  114.0 &  126.8&  115.0 \\
               &   3.5&  220.4&  199.3&  220.7&  203.7 &  207.5&  208.4 \\
               &   4.0&  280.5&  282.1&  283.7&  289.1 &  284.3&  306.6 \\
\noalign{\smallskip}
\hline
\noalign{\smallskip}
               &   2.0 &  9.7 &  6.1 &  9.8&  6.3&  9.9&  7.0 \\
               &   2.5 &  28.6&  18.2 &  28.9&  19.0&  29.3&  21.0 \\
$T_{\rm eff} = 4500K$&   3.0 &  74.4 &  50.5 &  75.0&  52.3&  76.2&  57.6 \\
               &   3.5 &  149.6&  116.0 &  150.5&  119.0&  153.7&  130.5 \\
               &   4.0 &  223.8&  203.2 &  224.9&  208.0&  233.1&  226.0 \\
\noalign{\smallskip}
\hline
\noalign{\smallskip}
               & 2.0& 3.6 &  2.6&  3.7&  2.7&   3.9&  2.9  \\
               & 2.5& 11.0&  8.0&  11.4&  8.3&   12.0&  9.1  \\
$T_{\rm eff} = 5000K$& 3.0& 32.0&  23.6&  33.1&  24.4&   34.6&  26.8  \\
               & 3.5& 80.5&  62.7&  82.7&  64.7&   86.3&  70.7  \\
               & 4.0& 154.2&  134.2&  157.1&  137.6&  162.5  &  148.3  \\
\noalign{\smallskip}
\hline
\noalign{\smallskip}
               & 2.0 & 1.5&  1.3&   1.6&  1.3&   1.7 & 1.4 \\
               & 2.5 & 4.8&  4.1&   5.0&  4.2&   5.3 & 4.5 \\
               & 3.0 & 14.6&  12.5&   15.2&  12.8&   16.1 & 13.7 \\
$T_{\rm eff} = 5500K$& 3.5 & 41.0 &  35.7&   42.6&  36.4&   44.8 & 39.0 \\
               & 4.0 & 95.6&  87.3&   98.7&  89.0&   103.3 & 94.7 \\
\noalign{\smallskip}
\hline
\noalign{\smallskip}
               & 2.5 &  2.3&  2.3&  2.5&  2.4 &   2.6&  2.5 \\
               & 3.0 &  7.3&  7.1&  7.9&  7.4 &   8.2&  7.7 \\
$T_{\rm eff}=6000K$& 3.5 &  21.5&  20.9&  23.2&  22.0 &   24.1&  22.8 \\
               & 4.0 &  56.8&  56.0&  60.5&  58.4 &   62.9&  60.7\\
\noalign{\smallskip}
\hline
\hline
\end{tabular}
\end{table*}

\begin {figure}
\psfig {file=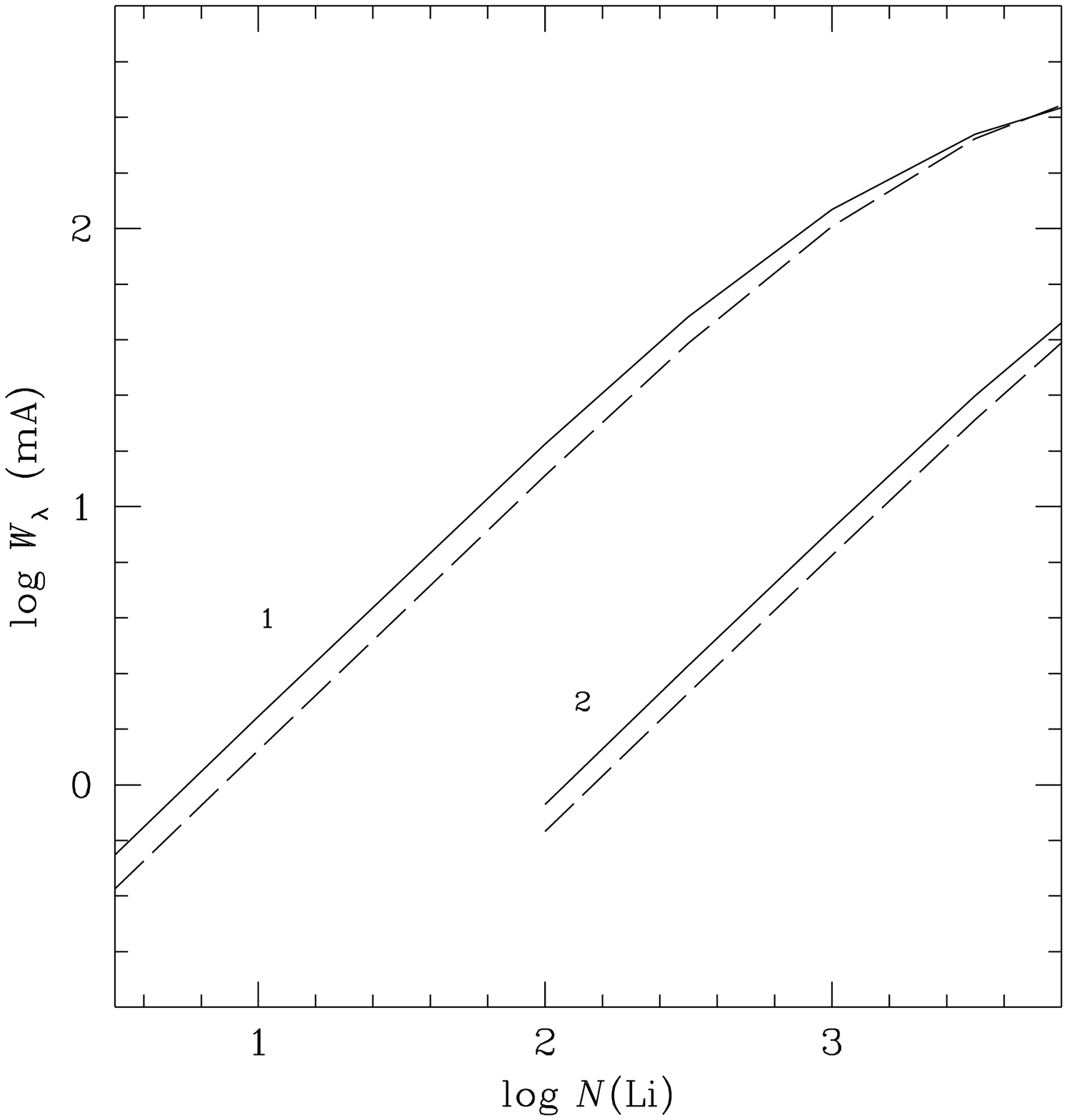,height=8.8cm}
\caption [] {The same as in Fig.~\ref{co67},  but for subordinate line at $\lambda 812.6$~nm}
\label {co81}
\end {figure}

The scarce dependence of all the considered \lii\ lines on $\lgg$  may
be explained in the following way. The statistical balance of lithium
in cool atmospheres is governed by the radiation field (Steenbock \&
Holweger 1984). The radiation field mean intensity  crucially
depends on the value of electron density ($N_{\rm e}$) in the stellar
photosphere. We found that differences of $N_{\rm e}$ in the region
$0.1 < \tau_{1\mu} < 1.0$ in the  photosphere of stars with different
$\lgg$'s are much more pronounced for $T_{\rm eff} \ge 4500$~K than for
lower $T_{\rm eff}$'s.  As a result, in the last case the intensity of
the outgoing radiation field  in the frequencies of lithium transitions
is less sensitive to $\lgg$. Moreover, the role of inelastic collisions
in the formation of statistical equilibrium decreases with lowering $T_{\rm
eff}$ due to the decrease of $N_{\rm e}$. Note that the independence
from $\lgg$ is seen only for unsaturated lines, which do not have
extended, sensitive to $\lgg$, wings.

\subsubsection {Comparison with other works}

The LTE and non-LTE curves of growth presented here agree well with
those used in the works of Mart\'\i n \etl\ (1994) and Garc\'\i a
L\'opez \etl\ (1994), the differences in equivalent width being  less
than 5 percent. There is a good agreement also  with Magazz\`u \etl's
(1992) results, except for cooler stars.  Reasons and consequences of
this disagreement at lower temperatures are discussed in Pavlenko
\etl\ (1995). Note that  in this paper we use a refined value of the
damping constant for the \lii\ resonance doublet.  We use also an
extended opacity list, that makes  the curves of growth presented here
more reliable, especially for lower effective temperatures. For these
points see also  Pavlenko \etl\ (1995).

\begin{table*}
\caption[]{Curves of growth of the $\lambda 812.6$ nm line}
\label{tab2}
\begin{tabular} {lrrrrrrr}
\hline
\hline
\noalign{\smallskip}
           &$\log N({\rm Li})$ & \multicolumn {6}{c}{$W_\lambda~({\rm mA})$}      \\ \cline {3-8}
\multicolumn{2}{c} { } & \multicolumn{2}{c} {$\log g = 3.0$} & \multicolumn{2}{c}
{$\log g = 3.5$} &\multicolumn{2}{c} {$\log g = 4.5$} \\ 
     & & LTE   &  NLTE   &  LTE  &   NLTE   &    LTE    &  NLTE  \\
\noalign{\smallskip}
\hline
\noalign{\smallskip}
               &   0.5&  0.6 &  0.4 &  0.5 & 0.4&  0.4 & 0.4 \\
               &   1.0&  1.8 &  1.3 &   1.5 &  1.3 &   1.2 &  1.3 \\
               &   1.5&    5.5 & 4.1 &   4.7 &  4.1 &   3.7&  4.2 \\
$T_{\rm eff} = 3500K$&   2.0&  16.8 &  13.0 &   14.5 &  12.9 &   11.5&  12.9  \\
               &   2.5&  48.1 &  38.7 &   41.8 &  38.3 &   33.8 &  38.2  \\
               &   3.0&  117.3&  101.8 &   104.7&  100.3 &   87.2&  100.0 \\
               &   3.5&  218.3&  210.8 &   203.0&  208.3 &   179.2&  207.4  \\
               &   4.0&  314.3&  333.0 &   302.4&  330.9 &   284.5&  331.5  \\
\noalign{\smallskip}
\hline
\noalign{\smallskip}
               &   1.0 &  0.6 & 0.4 &  0.6 & 0.4 &  0.5& 0.4 \\
               &   1.5 &  2.1 & 1.3 &   2.0 &  1.4 &   1.6&  1.4 \\
               &   2.0 &  6.4 & 4.2 &   6.3 &  4.3 &   4.9&  4.4  \\
$T_{\rm eff} = 4000K$&   2.5 &  19.6 & 13.0 &   19.3 &  13.4 &   15.2&  13.6  \\
               &   3.0 &  55.3 & 38.9 &   54.6&  40.0 &   43.9&  40.6  \\
               &   3.5 &  131.1 & 103.2 &   129.8&  105.8 &   109.4&  107.3  \\
               &   4.0 &  232.1 & 214.3 &   230.5&  218.8 &   209.8&  222.9  \\
\noalign{\smallskip}
\hline
\noalign{\smallskip}
               &   2.0 &  2.2 & 1.4 &   2.2 &  1.5 &  2.2&  1.6  \\
               &   2.5 &  6.8 & 4.5 &   6.8&  4.7 &  6.9&  5.1  \\
$T_{\rm eff} = 4500K$&   3.0 &  20.6 & 14.0 &   20.7&  14.5 &  20.9 &  15.9  \\
               &   3.5 &  58.0 & 41.5 &   58.3&  43.0 &  58.7&  47.1  \\
               &   4.0 &  136.3 & 108.1 &   136.7&  111.7 &  137.9&  122.4  \\
\noalign{\smallskip}
\hline
\noalign{\smallskip}
               & 2.0& 0.8 & 0.6&  0.8& 0.6 &  0.8& 0.7  \\
               & 2.5&  2.5 &  1.9&   2.6 &  2.0 &   2.7&  2.1  \\
$T_{\rm eff} = 5000K$& 3.0&  7.7 &  5.9&   8.0 &  6.1&   8.3&  6.7  \\
               & 3.5&  23.2 &  18.2&   24.0&  18.8 &   25.0&  20.5  \\
               & 4.0&  64.1 &  52.6&   65.9&  54.4 &    68.4& 59.4  \\
\noalign{\smallskip}
\hline
\noalign{\smallskip}
               & 2.5 &  1.0 & 0.9&     1.1& 1.0 &  1.2&  1.0 \\
               & 3.0 &  3.2 &  3.0&     3.4&  3.1 &  3.6&  3.3 \\
$T_{\rm eff} = 5500K$& 3.5 &  10.0 &  9.3&     10.6&  9.6 &  11.3&  10.2 \\
               & 4.0 &  29.7 &  27.9&     31.4 &  28.8 &  33.1&  30.8 \\
\noalign{\smallskip}
\hline
\noalign{\smallskip}
               & 2.5&  0.5 & 0.5 &  0.5 & 0.6 &  0.6 & 0.6 \\
               & 3.0&   1.5 &  1.6 &   1.7 &  1.7 &   1.8&  1.8  \\
$T_{\rm eff} = 6000K$& 3.5&   4.8 &  5.1 &   5.4 &  5.5 &   5.6&  5.6  \\
               & 4.0&   14.8 &  15.7 &   16.5 &  16.8 &   17.0&  17.3  \\
\noalign{\smallskip}
\hline
\hline
\end{tabular}
\end{table*}

\begin {figure}
\psfig {file=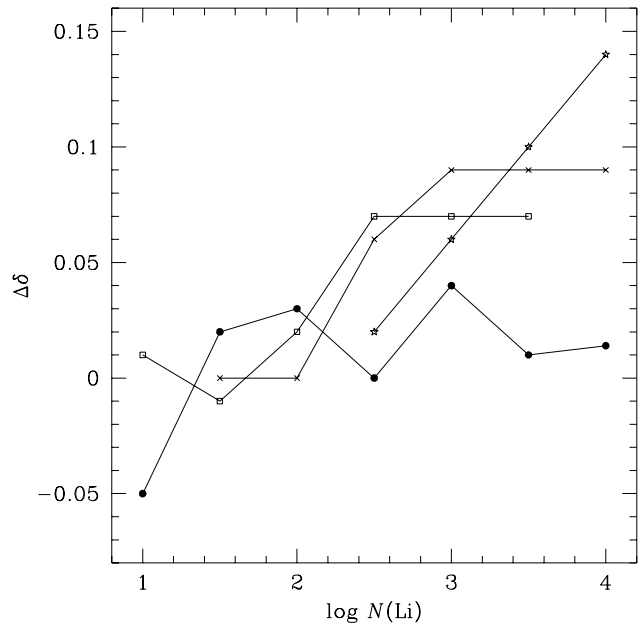,height=8.8cm}
\caption [] {The difference of non-LTE abundances corrections
$\Delta\delta = \Delta_{\rm us}$ - $\Delta_{\rm Car}$, computed by us
and Carlsson et al. (1994), respectively, for dwarf model atmospheres
4500/4.5 (dots), 5000/4.5 (squares), 5500/4.5 (crosses), 6000/4.5
(stars)}
\label {delta}
\end {figure}

Carlsson \etl\ (1994) did not give the computed equivalent widths of
lithium lines, but only non-LTE abundance corrections to apply to LTE
curves of growth for the lithium resonance doublet.  In order to
compare our results with theirs, we  use  the data of Table~1 to
compute our non-NLTE corrections. In Fig.~\ref{delta} we present
$\Delta\delta$, i.e.\ the difference between our corrections ($\Delta_{\rm
us}$) and those ($\Delta_{\rm Car}$) by Carlsson \etl\ (1994) for
several models in common.

We find that the agreement of these results is fairly good  (see also
Pavlenko 1995), the differences being less than 0.1~dex.  The greater
differences are seen in strong \lii\ lines and  are caused by the
cumulative effect of different model atmospheres, opacity sources,
cross sections, model atoms.  Note that the slope of curves of growth
for strong lithium lines decreases.  We point out that the differences
in the computed non-LTE abundance corrections are smaller than the
corrections themselves. We also note that the data for weak lines
formed deep in the atmosphere agree quite well. It is evident that
these lines are formed in  regions in the deep photosphere,  where the
temperature structures of models  by different authors are practically
the same.

Our referee, Dr. Carlsson,  kindly informed us about the results of a
comparison between our equivalent widths and those computed by Carlsson
et al. (1994) for the 5000/4.5 model atmosphere (solar metallicity).
For  LTE he found an agreement better than 10\% with a difference of
5\% for most abundances. The non-LTE equivalent widths agree within
15\%, but typically within 10\%.\@
  
\section {Conclusions}

In this paper we give, in tabular form, the  LTE \& non-LTE curves of
growth of lithium resonance ($\lambda 670.8$~nm) and subordinate lines
($\lambda 610.3$ and 812.6~nm)  computed using the newest Kurucz (1993)
models. These data may be used  for  lithium abundance determinations
as well as for the redetermination of  abundances obtained so far in the
frame of the LTE approach. In particular, we point out the importance
of the $\lambda 812.6$~nm data. This line lies in the red part of the
spectrum and its blending is not as strong as for the other two
lithium lines. The use of this line may give more accurate results.

Our theoretical curves of growth cover a different effective
temperature range than in  Carlsson \etl\ (1994). In fact, our main
interest is shifted towards lower temperatures which were not
considered by these authors.  We note a good agreement between our
results and those obtained in other works, despite the differences in
non-LTE procedure details.

Let us note that lithium in the stellar atmospheres  considered in
this paper exists mainly in ionized form. So the relatively small
deviations from   LTE obtained in our work for the coolest models of
our grid have not to be considered as an obvious result. Indeed, with
decreasing    effective temperature we have a  dramatic drop of the
electron pressure. On the other hand,  opacities in the frequencies of
lithium  transitions increase due to the strong molecular absorption.
In the case of high lithium abundances the resonance doublet becomes
very strong with saturated core and extended wings are formed deep in
the atmosphere, very close to LTE conditions. In the case of $T_{\rm
eff} < 3500$~K   lithium exists in stellar atmospheres mainly in the
form of neutral atoms, so non-LTE effects  should not be large a
priori. This was confirmed by direct computations by Pavlenko \etl\
(1995).

For the coolest stars of our grid ($3500 \le \Tf \le  4000$~K) we find
that equivalent widths and profiles of unsaturated \lii\ lines computed
in non-LTE  show only a very weak dependence on the $\lgg$ parameter.
Instead, we note that  the corresponding LTE curves of growth show a
quite strong dependence on gravity. The use of non-LTE curves of growth
gives a chance to minimize the possible effects of errors in $\lgg$  in
lithium abundance determinations using non-saturated absorption lines.
Moreover we note  that, using LTE calculations to perform Li abundance
determinations,  the knowledge of gravity seems not crucial for $\Tf
\ge 4500$~K, but it appears important for cooler stars. However, when
using non-LTE calculations, the reverse holds.

In this paper we have discussed results for classical model
atmospheres. The cool star spectra  often show emission lines created
in chromospheric-like features (CLF). However, as shown in Pavlenko
(1995) and Pavlenko \etl\ (1995), CLF  affect mainly   the LTE results
(both profiles and curves of growth). On the contrary, non-LTE results
show weak dependence on  CLF. Still in the case of the strongest CLF
producing {\it veiling}   the lithium lines may be severely affected.
Let us note that

\begin {itemize}

\item only the youngest and the most active stars may have such CLF producing additional continuum in the visible part of the spectrum;   

\item these results may be interpreted as a weak dependence of the (non-LTE) lithium lines on the structure of the outermost layers of model atmospheres.  At the present time  we cannot be sure that we describe these layers properly in the frame of the classical one-dimensional approach.

\end {itemize}

These two reasons give additional support to the use of non-LTE results
for numerical analysis of lithium lines in stellar spectra.

\acknowledgements
We thank Dr.  M.~Carlsson, our referee, for valuable comments which improved this paper.

\begin {thebibliography}{}

\item[]{} Auer, R.H., Heasley, I.N. 1976, ApJ, 205, 165

\item[]{} Carlsson, M., Rutten, R.J., Bruls, J.H.M.J., Shchukina, N.G. 1994,
A\&A, 288, 860

\item[]{} Crane, P.  1995, (Ed.) The Light Element Abundances, ESO Astrophysics Symposia, Springer

\item []{} Duncan, D.K. 1991, ApJ, 373, 250

\item []{} Drawin, H.W. 1968, Zeits.\ f.\ Physik, 211, 404 

\item []{} Drawin, H.W. 1969, Zeits.\ f.\ Physik, 225, 483

\item[]{} Garc\'\i a L\'opez, R.J., Rebolo, R., Mart\'\i n, E.L. 1994,
A\&A, 282, 518

\item[]{} Kurucz, R.L. 1979, ApJS, 40, 1

\item[]{} Kurucz, R.L. 1993, ATLAS9 Stellar Atmosphere Program and 2\,km~s$^{-1}$ grid (Kurucz CD-ROM No. 13)
 
\item[]{} Lambert, D.L. 1993, Physica Scripta, T47, 186

\item[]{} Magazz\`u, A., Rebolo, R., Pavlenko, Ya.V. 1992, ApJ, 392, 159

\item[]{} Magazz\`u, A., Mart\'\i n, E.L., Rebolo, R., Garc\'\i a L\'opez, R.J., Pavlenko, Ya.V. 1995, in The Light Element Abundances, ESO Astrophysics Symposia, Ed.\ P.~Crane, Springer, P. 280

\item[]{} Mart\'\i n, E.L., Rebolo, R., Magazz\`u, A., Pavlenko, Ya.V.
1994, A\&A, 282, 503

\item[]{} Pavlenko, Ya.V. 1984, Departure from LTE in atmospheres
of M-giants, Valgus, Tallinn, 1.

\item[]{} Pavlenko, Ya.V. 1989, Kinematika i Fizika Nebesnykh Tel, 5, 55

\item[]{} Pavlenko, Ya.V. 1992, SvA, 36 (6), 605

\item[]{} Pavlenko, Ya.V. 1994, Astrophys. Reports, 38, 531

\item[]{} Pavlenko, Ya.V. 1995, Mem.\ Soc.\ Astron.\ Ital., 
66, N. 2, p. 441

\item[]{} Pavlenko, Ya.V., Rebolo, R., Mart\'\i n, E.L., Garc\'\i a L\'opez, R.J. 1995, A\&A, 303, 807

\item[]{} Steenbock, W., Holweger, H. 1984, A\&A, 130, 319

\item[]{} Tsuji, T. 1994, in Molecules in the Stellar Environment, U.G. J{\o}rgensen (Ed.), Springer-Verlag, p. 79

\end {thebibliography}

\end {document}